\documentclass[prl,aps,twocolumn,superscriptaddress,
]{revtex4}
\usepackage{amsmath,amssymb,graphicx}
\usepackage{pxfonts}
\usepackage{graphicx}
\usepackage{color}
\usepackage{float}
\begin{document}


\title{Transport kinetics in excitonic Mott insulator}

\author{W.~J.~Brunner$^*$}
\affiliation{Department of Physics, University of California San Diego, La Jolla, CA 92093, USA}
\author{Zhiwen~Zhou$^*$}
\affiliation{Department of Physics, University of California San Diego, La Jolla, CA 92093, USA}
\author{Luc~M. Nguyen}
\affiliation{Department of Physics, University of California San Diego, La Jolla, CA 92093, USA}
\author{C.~A.~Lozano}
\affiliation{Department of Physics, University of California San Diego, La Jolla, CA 92093, USA}
\author{G.~Mercado}
\affiliation{Department of Physics, University of California San Diego, La Jolla, CA 92093, USA}
\author{L.~H.~Fowler-Gerace}
\affiliation{Department of Physics, University of California San Diego, La Jolla, CA 92093, USA}
\author{L.~V.~Butov} 
\affiliation{Department of Physics, University of California San Diego, La Jolla, CA 92093, USA}

\begin{abstract}
\noindent Indirect excitons (IXs) in a periodic moir{\'e} potential in a van der Waals heterostructure form an experimental platform for exploring Bose-Hubbard physics, 
including the superfluid state and the Mott insulator state. The realization of these collective states requires low-disorder materials. In particular, the localization due to the Mott insulator must be distinguished from disorder-induced localization. 
We studied the transport kinetics of IXs in a MoSe$_2$/WSe$_2$ van der Waals heterostructure 
with an anomalously high IX mobility, consistent with superfluidity. This high-mobility transport demonstrates that disorder-induced localization is negligible. We found a remarkable suppression of this efficient IX transport, directly indicating the Mott insulator.
As the IX density increases during the excitation pulse, the IX cloud initially expands and then contracts upon entering the Mott insulator phase. The emergence of IX cloud contraction is accompanied by a peak in the IX spatial distribution at the origin. Our simulations reproduce the observed non-monotonic IX transport kinetics, including the temporal expansion and contraction of the IX cloud, as well as the peak in its spatial distribution. 

\end{abstract}
\maketitle

The Bose-Hubbard (BH) model predicts a superfluid state when the number of bosons per site of a periodic potential is $N \sim 1/2$, and an insulating state, such as a Mott insulator or a Bose glass, when $N \sim 1$. In the latter case $N \sim 1$, all lattice sites are occupied and the repulsive interaction suppresses transport from site to site because hopping to an adjacent site requires overcoming the on-site interaction energy that makes the state insulating. In turn, transport without double occupancy makes the superfluid state more stable when $N \sim 1/2$~\cite{Fisher1989}.

A spatially indirect exciton (IX), also known as an interlayer exciton, is composed from an electron and a hole in separated layers~\cite{Lozovik1976}. The layer separation enables long IX lifetimes, which are significantly longer than those of regular direct excitons (DXs)~\cite{Zrenner1992}. The long lifetimes allow IXs to thermalize below the temperature of quantum degeneracy, creating an opportunity to realize quantum bosonic states~\cite{Keldysh1968, High2012}. 

A quantum system of IXs in a van der Waals heterostructure composed of atomically thin layers of transition metal dichalcogenides (TMDs) forms an experimental platform for exploring BH physics~\cite{Wu2018, Yu2018, Wu2017, Yu2017, Zhang2017a, Tang2020, Gu2022, Gotting2022, Xiong2023, Park2023, Fowler-Gerace2024, Zhou2024, Deng2025, Gao2024, Costa2026, Schutte2026}. Twisting TMD layers creates a periodic moir{\'e} potential for IXs. The periods and amplitudes of the moir{\'e} potentials provide the range of BH parameters -- the intersite hopping and repulsive on-site interaction -- that gives an opportunity to realize the superfluid and insulating phases predicted by the BH model with IXs. The repulsive interparticle interaction in this system originates from the out-of-plane dipoles of the IXs~\cite{Yoshioka1990} as confirmed by the studies of IX interaction in TMD heterostructures~\cite{Rivera2015, Nagler2017, Kremser2020, Sun2022}. IX filling of the moir{\'e} lattice sites can be controlled by an optical excitation that generates IXs. Compared to atoms in optical lattices~\cite{Greiner2002} and IXs in patterned-gate lattices in GaAs heterostructures~\cite{Remeika2012, Lagoin2022}, IXs in moir{\'e} lattices in TMD heterostructures feature strong intersite hopping and on-site interaction, enabling the creation of high-temperature collective states in periodic potentials~\cite{Capogrosso-Sansone2008}. 

IX transport in TMD heterostructures is intensively studied both by cw~\cite{Rivera2016, Jauregui2019, Unuchek2019a, Unuchek2019, Liu2019, Choi2020, Huang2020, Shanks2022, Rossi2023, Gao2024, Zhang2024, Fowler-Gerace2021, Peng2022, Troue2023, Fowler-Gerace2024, Zhou2024, Cutshall2025} and time-resolved~\cite{Yuan2020, Li2021, Wang2021, Sun2022, Tagarelli2023, Wietek2024, Deng2025, Zhou2026} imaging of IX photoluminescence (PL) cloud expansion. 
Due to their long lifetimes, delocalized IXs can travel over long distances. However, 
disordered potentials, including disordered moir{\'e} potentials, localize IXs and suppress IX transport~\cite{Rivera2016, Jauregui2019, Unuchek2019a, Unuchek2019, Liu2019, Choi2020, Huang2020, Shanks2022, Rossi2023, Gao2024, Zhang2024, Fowler-Gerace2021, Peng2022, Troue2023, Fowler-Gerace2024, Zhou2024, Cutshall2025, Yuan2020, Li2021, Wang2021, Sun2022, Tagarelli2023, Wietek2024, Deng2025, Zhou2026}. 
In contrast, the realization of collective states, such as the superfluid and the Mott insulator, 
require low-disorder materials. Low disorder is crucial for the superfluid, as disorder suppresses this state. Low disorder is also crucial for the Mott insulator, as the localization due to this collective state must be distinguished from disorder-induced localization. 

Earlier studies revealed a high-energy IX PL line appearing at high densities~\cite{Xiong2023, Park2023, Zhou2024, Deng2025}. This line indicates that two IXs occupy a moir{\'e} cell. However, the appearance of this line does not prove a Mott insulator state, because establishing an insulator state requires transport measurements that show a suppression of transport. Mere transport suppression also does not prove the Mott insulator, because transport can be suppressed for other reasons, such as disorder, as noted above. Detecting the Mott insulator requires transport measurements, which show highly efficient transport in the superfluid regime at a moir{\'e} cell occupation of $N \sim 1/2$ and the suppression of this transport in the Mott insulator at $N \sim 1$.

In this work, we studied the transport kinetics of IXs in a MoSe$_2$/WSe$_2$ van der Waals heterostructure 
with an anomalously high IX mobility, consistent with superfluidity, at $N \sim 1/2$. This high-mobility IX transport demonstrates a regime where disorder-induced localization becomes negligible, indicating that the requirements for the realization of the Mott insulator are met in the heterostructure. We found a remarkable suppression of this efficient IX transport, directly indicating the Mott insulator at $N \sim 1$. As the IX density increases during the excitation pulse, the IX cloud initially expands and then contracts upon entering the Mott insulator phase.

\begin{figure}
\begin{center}
\includegraphics[width=8.7cm]{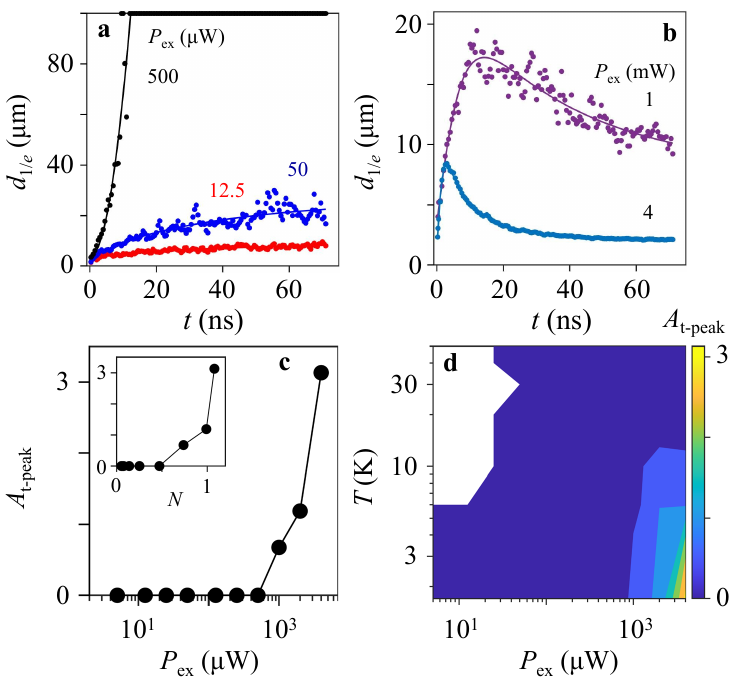}
\caption{Non-monotonic IX transport kinetics.
(a) Monotonic IX transport kinetics $d_{1/e}(t)$ for excitation density $P_{\rm ex} \sim 12.5,$ 50, and 500~$\mu$W 
($N \sim 0.06,$ 0.08, and 0.5). The heterostructure size allows determining $d_{1/e}$ up to $\sim 100$~$\mu$m~\cite{Zhou2024} and the data with higher $d_{1/e}$ are presented by points on the edge. 
(b) Non-monotonic IX transport kinetics $d_{1/e}(t)$ for $P_{\rm ex} \sim 1$ and 4~mW ($N \sim 0.7$ and 1). The non-monotonic IX transport kinetics is characterized by a peak in $d_{1/e}(t)$. 
(c,d) The parameter dependence of the peak contrast $A_{\rm t-peak} = d^{\rm m}_{1/e} / d^{\rm l}_{1/e} - 1$, where $d^{\rm m}_{1/e}$ corresponds to the peak maximum and $d^{\rm l}_{1/e}$ to long delay times.  
The values of $A_{\rm t-peak}$ in (c,d) are obtained using smoothed data, shown by lines in (a,b), and $t \sim 70$~ns for $d^{\rm l}_{1/e}$.
(c): $A_{\rm t-peak}$ vs. $P_{\rm ex}$. Insert: $A_{\rm t-peak}$ vs. $N$. 
(d): $A_{\rm t-peak}$ vs. $P_{\rm ex}$ and $T$.
$t = 0$ corresponds to the excitation pulse start. $T = 1.7$~K in (a-c). $N$ values are estimated within the mean field model~\cite{Yoshioka1990} as outlined in the text. 
}
\end{center}
\label{fig:spectra}
\end{figure}

\begin{figure}
\begin{center}
\includegraphics[width=8.7cm]{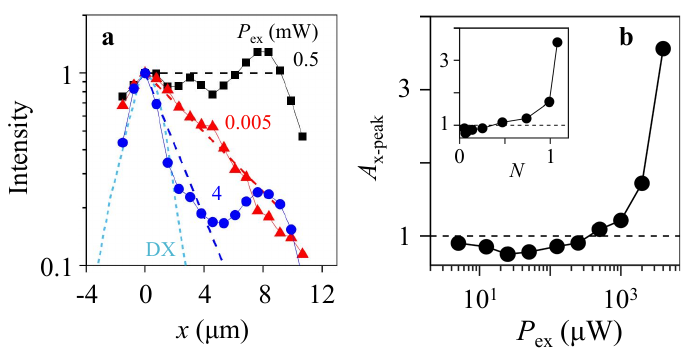}
\caption{Spatial distribution of IXs: The peak at the origin.
(a) IX signal vs. the distance $x$ from the excitation spot center, $I(x)$, for $P_{\rm ex} \sim 0.005$, 0.5, and 4~mW ($N \sim 0.05$, 0.5, and 1). The excitation spot is shown by the cyan dotted line, which presents the DX PL profile in the MoSe$_2$ monolayer and is close to the laser profile for short-range DX transport. The dashed lines show least-squares fits of $I(x)$ to exponential decays in the region from the excitation spot to the HS edge, $x = 0 - 9$~$\mu$m. The slopes of the lines are used to estimate $d_{1/e}$. For $N \sim 1$, a peak in $I(x)$ emerges at the origin of the IX cloud. This peak is quantified by the contrast $A_{\rm x-peak} = (I_0/I_3)/(I_3/I_6)$, where $I_0$, $I_3$, and $I_6$ is IX signal at $x = 0$, 3, and 6~$\mu$m, respectively.
(b) $A_{\rm x-peak}$ vs. $P_{\rm ex}$. Insert: $A_{\rm x-peak}$ vs. $N$.  
$I(x)$ in (a) are normalized to the same value at $x = 0$. $N$ values are estimated within the mean field model~\cite{Yoshioka1990} as outlined in the text.
}
\end{center}
\label{fig:spectra}
\end{figure}

\vskip5mm
{\bf Experiments}

In the MoSe$_2$/WSe$_2$ van der Waals heterostructure, the adjacent MoSe$_2$ monolayer and WSe$_2$ monolayer form separated electron and hole layers for IXs~\cite{Rivera2015}. The heterostructure details are presented in Supplementary Information (SI).

IXs are generated by a focused laser excitation with the laser excitation energy resonant to DX in WSe$_2$ heterostructure layer. The laser pulses have step-like rectangular shape with an 84 ns pulse width and 0.1 (0.3) ns sharpness of the front (back) edge. The IX transport is measured by time- and spatially-resolved IX PL imaging. The measurement details are presented in SI.

The IX cloud expands due to IX transport from the excitation spot. The extension of the IX cloud is described by its $1/e$ decay distance, $d_{1/e}$. The temporal kinetics of $d_{1/e}(t)$ characterize IX transport. 

Figure~1a shows the IX transport kinetics $d_{1/e}(t)$ for different laser excitation powers $P_{\rm ex}$. The IX density grows from the start of laser excitation and saturates with increasing delay time. At sufficiently long delay times, both the IX density and IX transport distance saturate at the values corresponding to cw laser excitation (as outlined in SI). 

The IX density in the saturation, or cw, regime is estimated within the mean-field model~\cite{Yoshioka1990} from the IX energy shift $\delta E = n u_0$, $u_0 = 4 \pi e^2 d_z/\varepsilon$, where $d_z$ is the separation between the electron and hole layers and $\varepsilon$ is the dielectric constant ($d_z \sim 0.6$~nm and $\varepsilon \sim 7.4$ for the MoSe$_2$/WSe$_2$ heterostructure~\cite{Laturia2018}). The moir{\'e} superlattice period $b \sim 17$~nm in the heterostructure (outlined in SI) and the IX density give the estimated IX filling of the moir{\'e} lattice sites. $N = 1$ corresponds to $n \sim 4 \cdot 10^{11}$~cm$^{-2}$ and $n u_0 \sim 6$~meV~\cite{Yoshioka1990} for the heterostructure. The mean-field model~\cite{Yoshioka1990} is commonly used to estimate IX densities in various heterostructures, e.g. in Refs.~\cite{Larionov2000, Ivanov2002, Butov2002, Vogele2009, Alloing2012, Szwed2024, Jauregui2019, Unuchek2019, Sun2022, Zhang2024, Fowler-Gerace2021, Fowler-Gerace2024}, and is applied in this work (as outlined in SI). We note that including correlations lowers $u_0$ in comparison to the mean-field value that leads to a higher estimated $n$ for a given $\delta E$~\cite{Schindler2008, Remeika2009, Laikhtman2009, Ivanov2010, Remeika2015, Steinhoff2025}. Estimates based on the mean-field model~\cite{Yoshioka1990} are more accurate for higher densities~\cite{Remeika2015}.

Figure~1a shows that for lower $P_{\rm ex}$ and, in turn, lower $N$, the IX cloud monotonically expands with time. For $N$ well below 1/2, $d_{1/e}(t)$ increases slowly, consistent with diffusive transport in a disordered in-plane potential. Diffusive IX transport kinetics have been intensively studied in TMD heterostructures~\cite{Yuan2020, Li2021, Wang2021, Sun2022, Tagarelli2023, Wietek2024, Zhou2026}. Efficient IX transport kinetics, characterized by a rapid increase in $d_{1/e}(t)$ and high IX mobility, is observed when $N$ approaches 1/2 (Fig.~1a). The regime of efficient IX transport with anomalously high IX mobility, which is consistent with a superfluid phase, was studied in Ref.~\cite{Zhou2026} and outlined in the introduction. 

A monotonic increase of $d_{1/e}(t)$ over time characterizes both diffusive transport for $N << 1/2$ and superfluid transport for $N \sim 1/2$ (Fig.~1a). However, a non-monotonic kinetics of $d_{1/e}(t)$ with time is observed when $N$ approaches 1: the IX cloud initially expands and then contracts (Fig.~1b).  

The non-monotonic IX transport kinetics is characterized by a peak in $d_{1/e}(t)$ (Fig.~1b). The peak contrast is quantified by $A_{\rm t-peak} = d^{\rm m}_{1/e} / d^{\rm l}_{1/e} - 1$, where $d^{\rm m}_{1/e}$ corresponds to the peak maximum and $d^{\rm l}_{1/e}$ to long delay times where $d_{1/e}$ approaches its cw value. The values of $A_{\rm t-peak}$ in Fig.~1c,d are obtained using smoothed data, shown by lines in Fig.~1a,b, and $t \sim 70$~ns for $d^{\rm l}_{1/e}$. Figure~1c shows that the peak in $d_{1/e}(t)$ emerges when $N$ approaches 1. The peak in $d_{1/e}(t)$ vanishes above $T \sim 10$~K (Fig.~1d). The non-monotonic IX transport kinetics (Fig.~1b) observed upon entering the Mott insulator regime is compared with simulations and discussed in the following sections. 

The emergence of the non-monotonic temporal kinetics of IX transport upon entering the Mott insulator regime (Fig.~1) is accompanied by a peak in the spatial distribution of IXs at the origin (Fig.~2). For lower $P_{\rm ex}$ corresponding to $N \lesssim 1/2$, the IX signal $I(x)$ nearly exponentially varies with distance $x$ from the excitation spot center; In contrast, for higher $P_{\rm ex}$, corresponding to $N \sim 1$, a peak in $I(x)$ emerges at the IX cloud origin (Fig.~2a). This peak is quantified by the non-exponentiality contrast $A_{\rm x-peak} = (I_0/I_3)/(I_3/I_6)$, where $I_0$, $I_3$, and $I_6$ is IX signal at $x = 0$, 3, and 6~$\mu$m, respectively. For an exponential decay of $I(x)$, $A_{\rm x-peak} = 1$. Consequently, for a cloud that originates at $x = 0$ and decays exponentially with increasing $x$, $A_{\rm x-peak}$ is less than 1 because of the smooth $I(x)$ maximum at the origin. Nearly exponential spatial decays of the IX signal with $A_{\rm x-peak}$ less than 1 are observed for $N \lesssim 1/2$ (Fig.~2b). In contrast, a peak in IX spatial distribution, with $A_{\rm x-peak}$ exceeding 1, emerges at the origin upon entering the Mott insulator regime, $N \sim 1$, (Fig.~2b). (We note signal modulation along the IX path; this modulation is related to sample inhomogeneities, and its effect on the IX transport considered in this work is insignificant.) The peak at the origin in the IX spatial distribution (Fig.~2) observed upon entering the Mott insulator regime is compared with simulations and discussed below.

\begin{figure}
\begin{center}
\includegraphics[width=8.7cm]{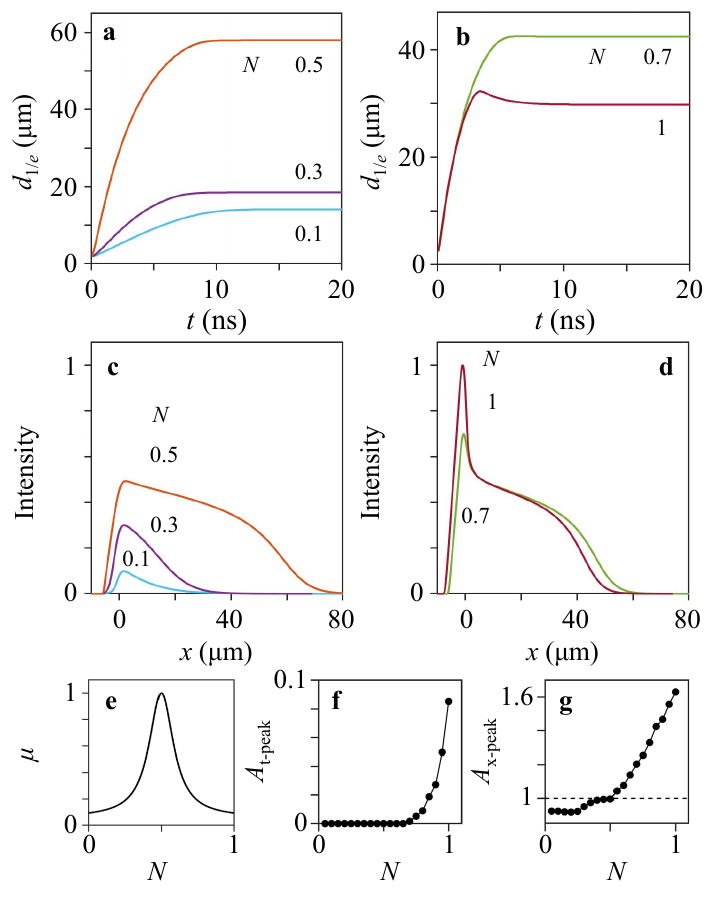}
\caption{Simulated IX transport kinetics and spatial distribution.
(a,b) IX transport kinetics $d_{1/e}(t)$ for $N = 0.1$, 0.3, and 0.5 (a), and $N = 0.7$ and 1 (b).
Non-monotonic IX transport kinetics with a peak in $d_{1/e}(t)$ emerges for $N \sim 1$ (b).
(c,d) Spatial profiles of IX signal $I(x)$ for $N = 0.1$, 0.3, and 0.5 (c), and $N = 0.7$ and 1 (d).
A peak in $I(x)$ emerges at the origin for $N \sim 1$. 
(e) IX mobility $\mu(N)$ in the simulations. $\mu(1/2) = 1.7 \cdot 10^5$~cm$^2$/eVs.
(f) Contrast of the temporal peak $A_{\rm t-peak}$. 
(g) Contrast of the spatial peak $A_{\rm x-peak}$. 
For all data, $T = 1.7$~K, the spatial and temporal profiles of the laser excitation correspond to the experiment. 
}
\end{center}
\label{fig:spectra}
\end{figure}

\vskip5mm
{\bf Simulations}

As outlined in the introduction, the BH model predicts a superfluid state for $N \sim 1/2$, and an insulating state, such as a Mott insulator or a Bose glass, for $N \sim 1$~\cite{Fisher1989}. For low $N$, unscreened disorder contributes to IX localization, and IX mobility increases with density due to screening of the disorder potential~\cite{Ivanov2002, Remeika2009}. The IX transport enhancement with $N$ followed by its suppression for $N \sim 1$, predicted by the BH model, is simulated by drift-diffusion transport with the IX mobility $\mu (N)$ first increasing with $N$, reaching maximum at $N = 1/2$, and falling at $N = 1$ (Fig.~3e). These proof-of-principle simulations do not aim to provide a precise description of transport in either the IX superfluid or Mott insulator regimes. However, they show that the non-monotonic dependence of IX transport on the moir{\'e} cell occupancy qualitatively reproduces the observed kinetics, including the temporal expansion and contraction of the IX cloud (Fig.~1), as well as the peak at the origin in the IX spatial distribution (Fig.~2), upon entering the Mott insulator regime.

The transport kinetics is simulated by the drift-diffusion equation for IX density $n$
\begin{eqnarray}
\frac{\partial n}{\partial t} = \nabla \left[ D \nabla n + \mu n \nabla (n u_0 + U) \right] + \Lambda - \frac{n}{\tau}\,
\end{eqnarray}
The first term describes IX diffusion, the second and the third IX drift due to interaction and in-plane potential energy landscape, and the last two terms IX generation by laser excitation and IX recombination~\cite{Ivanov2002}. The in-plane potential energy landscape, $U$, arises from variations of the local IX environment in the heterostructure and is outlined in SI. We use the Einstein relation for IX diffusivity $D = \mu k_B T$ ($k_{\rm B}$ is the Boltzmann constant) and density-dependent IX mobility $\mu = \mu(N)$ shown in Fig.~3e. The IX lifetime $\tau$ was measured in Ref.~\cite{Zhou2026}. The spatial profile of the IX generation, $\Lambda(x)$, corresponds to the laser excitation profile (dotted line in Fig.~2a), and its temporal profile matches the laser pulses outlined above. The signal intensity is approximated as proportional to IX density, which consequently reflects the IX filling of the moir{\'e} lattice sites. The latter is plotted in Fig.~3c,d. For clarity, we use the notation $N$ for the moir{\'e} cell occupancy at the origin at sufficiently long delay times when it saturates at the value corresponding to cw laser excitation, as outlined in the Experiments section. For $\mu(N)$ (Fig.~3e), $N$ refers to the moir{\'e} cell occupancy at any given time and position. The time-integrated signal in Figs.~3c,d is similar to the signal at long delay times.

The simulations qualitatively reproduce the features observed in the experiment: (i) Both the experiment (Fig.~1a) and simulations (Fig.~3a) show monotonic IX transport kinetics for $N \lesssim 1/2$. However, both the experiment (Fig.~1b,c) and simulations (Fig.~3b,f) show non-monotonic transport kinetics with the IX cloud initially expanding and then contracting upon entering the Mott insulator regime. (ii) Both the experiment (Fig.~2) and simulations (Fig.~3c,d,g) show that the emergence of the non-monotonic transport kinetics is accompanied by a peak at the origin in the IX spatial distribution. The data are discussed below.

\vskip5mm
{\bf Discussion}

A monotonic increase of the transport decay distance $d_{1/e}(t)$ with time is typical for IX transport in the excitonic regime, as measured both in TMD~\cite{Yuan2020, Li2021, Wang2021, Sun2022, Tagarelli2023, Wietek2024, Zhou2026} and GaAs~\cite{Voros2005, Gartner2006, Vogele2009, Dorow2018} heterostructures. It is consistent with diffusion, drift, or superfluid transport with a constant diffusivity, mobility, or velocity, respectively~\cite{Yuan2020, Li2021, Wang2021, Sun2022, Tagarelli2023, Wietek2024, Zhou2026, Voros2005, Gartner2006, Vogele2009, Dorow2018}. A monotonic increase of $d_{1/e}(t)$ (Fig.~1a) is also consistent with IX transport, which increases with density. This behavior is modelled by $\mu(N)$ increasing with $N$ in the range $0 \le N \le 1/2$ (Fig.~3e) and is shown in Fig.~3a. 

The observed non-monotonic IX transport kinetics, characterized by the expansion followed by contraction of the IX cloud (Fig.~1b), corresponds to a non-monotonic variation of IX transport with density. The simulations with a non-monotonic variation of IX mobility with density, $\mu(N)$ (Fig.~3e), show non-monotonic IX transport kinetics, with an increase of $d_{1/e}(t)$ followed by its decrease over time (Fig.~3b), in qualitative agreement with the experiment (Fig.~1b). 

A reduction of IX mobility as density increases is inconsistent with diffusive transport, for which mobility increases with density due to IX screening of the disorder potential~\cite{Ivanov2002, Remeika2009}. Furthermore, IX transport increases with density due to enhanced interaction-induced drift from the origin (as described by the second term in Eq.~1). 

The suppression of IX transport as density increases, modelled by a reduction of IX mobility at high $N$ (Fig.~3e), is consistent with the BH model, which predicts transport suppression upon entering the Mott insulator regime, $N \sim 1$. Both in the experiment (Fig.~1) and in the simulations (Fig.~3a,b,f), transport suppression emerges upon entering the Mott insulator regime, consistent with the BH model.  

As found in earlier studies~\cite{Zhou2026}, IX superfluidity, characterized by the rapid expansion of the IX cloud and anomalously high IX mobility, is suppressed at $T \gtrsim 10$~K. This is qualitatively consistent with the estimated superfluidity temperature in the heterostructure moir{\'e} potential~\cite{Capogrosso-Sansone2008} as outlined in Ref.~\cite{Zhou2026}. A slower IX cloud expansion at high temperatures prevents the system from achieving high $d_{1/e}(t)$ values and, in turn, suppresses the development of a peak in $d_{1/e}(t)$. A peak in $d_{1/e}(t)$ does not appear at high temperatures (Fig.~1d).

The peak at the origin in the IX spatial distribution observed in the experiment (Fig.~2) and in the simulations (Fig.~3d,g) is consistent with the IX transport suppression upon entering the Mott insulator regime. When $N$ approaches 1 and IX transport in the Mott insulator phase is suppressed, IXs accumulate at the origin, creating the peak in the spatial distribution. As IXs travel away from the origin, the moir{\'e} cell occupancy decreases. Consequently, IX transport initially increases while occupancy remains above 1/2, then reduces once occupancy drops below 1/2, as modelled by $\mu(N)$ in Fig.~3e. As a result, the IX peak at the origin is followed by a gradual density decrease with increasing $x$ in the fast transport region, and then by a rapid decline in the slow transport region (Fig.~3d). This latter decline occurs at high $x$, beyond the heterostructure edge in the simulations (Fig.~3d).

In summary, we studied the kinetics of IX transport in the excitonic Mott insulator regime in a van der Waals heterostructure. We found anomalous IX transport kinetics: as the IX density increases during the excitation pulse, the IX cloud initially expands and then contracts upon entering the Mott insulator phase. The emergence of IX cloud contraction is accompanied by a peak in the IX spatial distribution at the origin. Our simulations reproduce the observed non-monotonic IX transport kinetics, including the temporal expansion and contraction of the IX cloud, as well as the peak in its spatial distribution. 

\vskip5mm
{\bf Acknowledgements} 

We thank M.~M.~Fogler, D.~E.~Parker, and A.~H. MacDonald for discussions and A.~K.~Geim for teaching us manufacturing TMD heterostructures. The studies were supported by the Department of Energy, Office of Basic Energy Sciences, under award DE-FG02-07ER46449. The heterostructure manufacturing was supported by NSF Grant 1905478. The data analysis was supported by NSF Grant 2516006.

\vskip3mm
$^*$ These authors contributed equally: W.~J.~Brunner, Zhiwen~Zhou.

\end{document}



\title{Supporting Information for\\
Transport kinetics in excitonic Mott insulator}

\author{W.~J.~Brunner$^*$}
\affiliation{Department of Physics, University of California San Diego, La Jolla, CA 92093, USA}
\author{Zhiwen~Zhou$^*$}
\affiliation{Department of Physics, University of California San Diego, La Jolla, CA 92093, USA}
\author{Luc~M. Nguyen}
\affiliation{Department of Physics, University of California San Diego, La Jolla, CA 92093, USA}
\author{C.~Lozano}
\affiliation{Department of Physics, University of California San Diego, La Jolla, CA 92093, USA}
\author{G.~Mercado}
\affiliation{Department of Physics, University of California San Diego, La Jolla, CA 92093, USA}
\author{L.~H.~Fowler-Gerace}
\affiliation{Department of Physics, University of California San Diego, La Jolla, CA 92093, USA}
\author{L.~V.~Butov} 
\affiliation{Department of Physics, University of California San Diego, La Jolla, CA 92093, USA}

\begin{abstract}
\end{abstract}
\maketitle

\renewcommand*{\thefigure}{S\arabic{figure}}

{\bf Heterostructure}

The van der Waals MoSe$_2$/WSe$_2$ heterostructure (Fig.~S1a) was assembled using the dry-transfer peel technique~\cite{Withers2015}, as described in Ref.~\cite{Fowler-Gerace2024} where the same heterostructure was used for cw studies of IXs. The thickness of the bottom and top hBN layers is $\sim 40$~nm and $\sim 30$~nm, respectively. The MoSe$_2$ monolayer is on top of the WSe$_2$ monolayer. A rotational alignment between the WSe$_2$ and MoSe$_2$ monolayers was performed using long WSe$_2$ and MoSe$_2$ edges (Fig.~S1b). The twist angle between the monolayers, $\delta \theta = 1.1^\circ$, corresponding to the moir{\'e} superlattice period $b = 17$~nm, agrees with the angle between MoSe$_2$ and WSe$_2$ edges in the heterostructure, as outlined in Ref.~\cite{Fowler-Gerace2024}. The accuracies of estimating $\delta \theta$ using long WSe$_2$ and MoSe$_2$ edges is comparable to that using SHG. We do not use the latter for additional estimates of $\delta \theta$ since the intense excitation pulses in SHG measurements may cause deterioration of the heterostructure and suppress IX transport. Moir{\'e} potentials can be affected by atomic reconstruction~\cite{Weston2020, Rosenberger2020, Zhao2023} and disorder and may vary over the heterostructure area. The IX $g$-factor $g \sim - 16$ measured in the MoSe$_2$/WSe$_2$ heterostructure corresponds to H stacking~\cite{Seyler2019, Wozniak2020}.

A microscope image (Fig.~S1b) shows the layer boundaries of the heterostructure. The hBN layers cover the entire areas of MoSe$_2$ and WSe$_2$ layers. A narrow multilayer graphene electrode, initially placed on top of the heterostructure around $x = 2$~$\mu$m for $y = 0$, was detached. 

As outlined in the main text, disordered potentials localize IXs and suppress IX transport. Therefore, a heterostructure with sufficiently low disorder is needed for the realization of collective states, such as superfluid and Mott insulator states. IX transport with high IX mobility was recently observed in the MoSe$_2$/WSe$_2$ heterostructure~\cite{Zhou2026}. The high IX mobility in the presence of heterostructure disorder, as well as the range of parameters over which it is observed, are qualitatively consistent with the Bose-Hubbard model predictions for a superfluid state, as described in Ref.~\cite{Zhou2026} and outlined in the main text. 
This indicates that the requirements of low disorder for the realization of the Mott insulator are met in the heterostructure.
Therefore, we study IX transport in the Mott insulator regime in the same heterostructure. This work demonstrates non-monotonic temporal dynamics of the IX cloud in the Mott insulator regime. It is essential to study this phenomenon in other samples with different heterostructure parameters. Manufacturing samples with different parameters, all with sufficiently small disorder, is challenging. This is the subject for future work.

\begin{figure}
\begin{center}
\includegraphics[width=8.5cm]{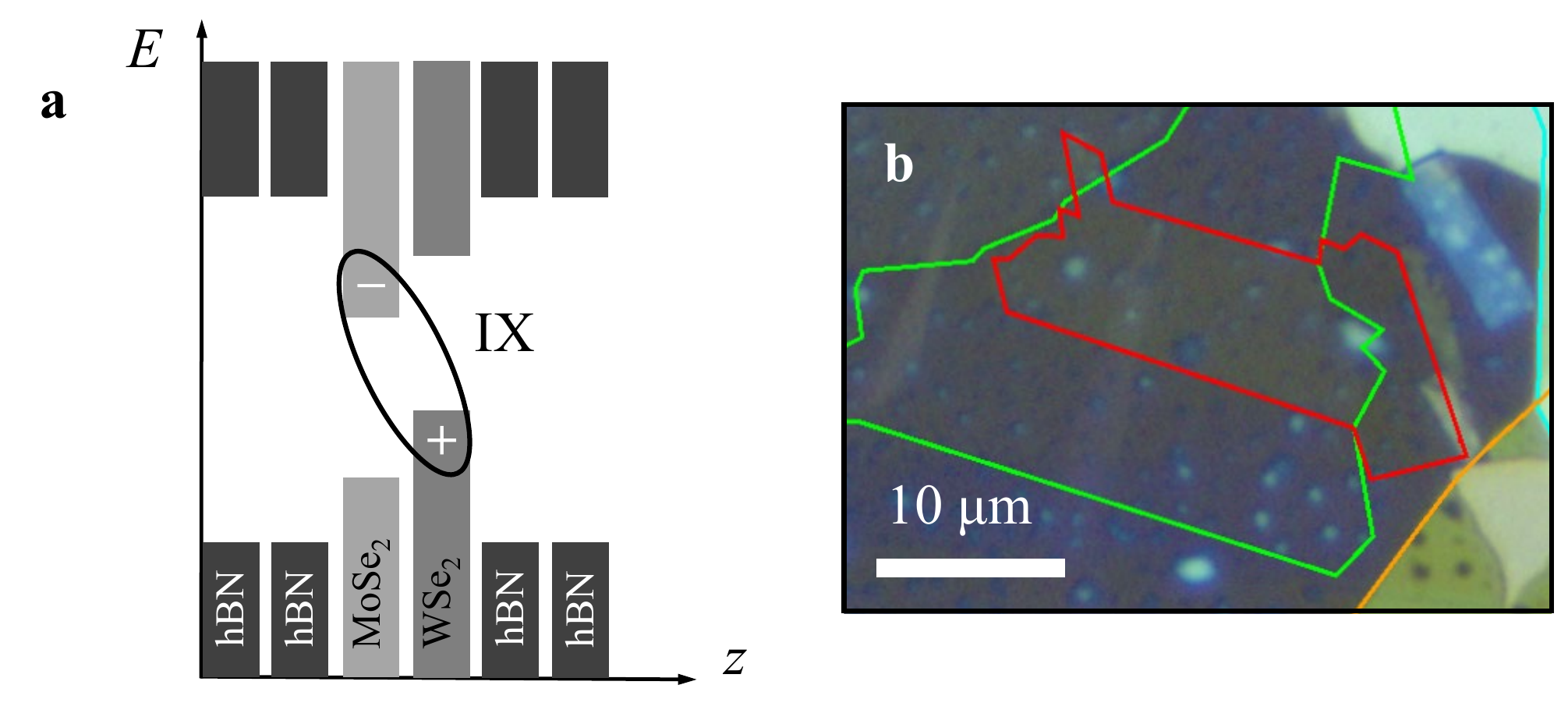}
\caption{(a) Energy-band diagram of the MoSe$_2$/WSe$_2$ heterostructure. The oval indicates IX composed of an electron (--) and a hole (+).
(b) A microscope image showing the layer boundaries of the heterostructure. The green, red, cyan, and orange lines indicate the boundaries of the WSe$_2$ monolayer, the MoSe$_2$ monolayer, the bottom hBN layer, and the top hBN layer, respectively. 
}
\end{center}
\label{fig:spectra}
\end{figure}

\vskip5mm
{\bf Measurements}

Excitons were generated by laser excitation focused onto a spot with a half-width at half-maximum $\sim 1$~$\mu$m, using an excitation energy $E_{\rm ex} = 1.689$~eV resonant with DXs in the WSe$_2$ heterostructure layer. The excitation spot profile is shown in Fig.~2a in the main text. The laser pulses had step-like rectangular shape with an 84 ns pulse width, 0.1 (0.3) ns sharpness of the front (back) edge (the pulses rise and fall $e$ times within 0.1 and 0.3~ns, respectively), and 300~ns pulse period. The $\sim 220$~ns off time significantly exceeded the few-ns IX lifetime and was sufficient for a substantial decay of the IX signal.

The IX transport was measured by time- and spatially-resolved optical imaging. An avalanche photodiode (APD) was moved in the optical image plane to measure the IX PL kinetics $I(x, t)$. An image magnification of x92 and an APD window diameter of 100~$\mu$m provided a spatial integration of $\sim 1.1$~$\mu$m for the imaging setup. The time resolution was 0.1~ns. Both the spatial and temporal resolutions of the setup were sufficient for imaging the IX transport. The IX emission in the transport kinetics measurements was spectrally selected using a long-pass filter with a cut-off wavelength of 850~nm (1.459~eV). 

The $1/e$ decay distance, $d_{1/e}$, was used to quantify the spatial extension of the IX cloud, $I(x, t)$, as outlined in the main text. Alternatively, the mean-squared displacement (MSD) can be used to quantify $I(x, t)$. MSD$^{1/2}$ is close to $d_{1/e}$, as verified in Ref.~\cite{Zhou2026}. However, $d_{1/e}$ is more accurate for describing experimental data. In particular, $d_{1/e}$ accurately quantifies the expansion of spatial profiles $I(x, t)$ even when $d_{1/e}$ exceeds the sample size, whereas MSD is affected by the finite sample size and cannot be used in this case. 

PL spectra were measured using a spectrometer with 0.2~meV resolution and a liquid-nitrogen-cooled CCD. The experiments were performed in a 4He cryostat. The sample was mounted on an Attocube $xyz$ piezo translation stage, providing adjustment of the sample position relative to a focusing lens inside the cryostat. All phenomena presented in this work are reproducible after multiple cooling cycles from room temperature to 2~K.

\begin{figure}
\begin{center}
\includegraphics[width=8.5cm]{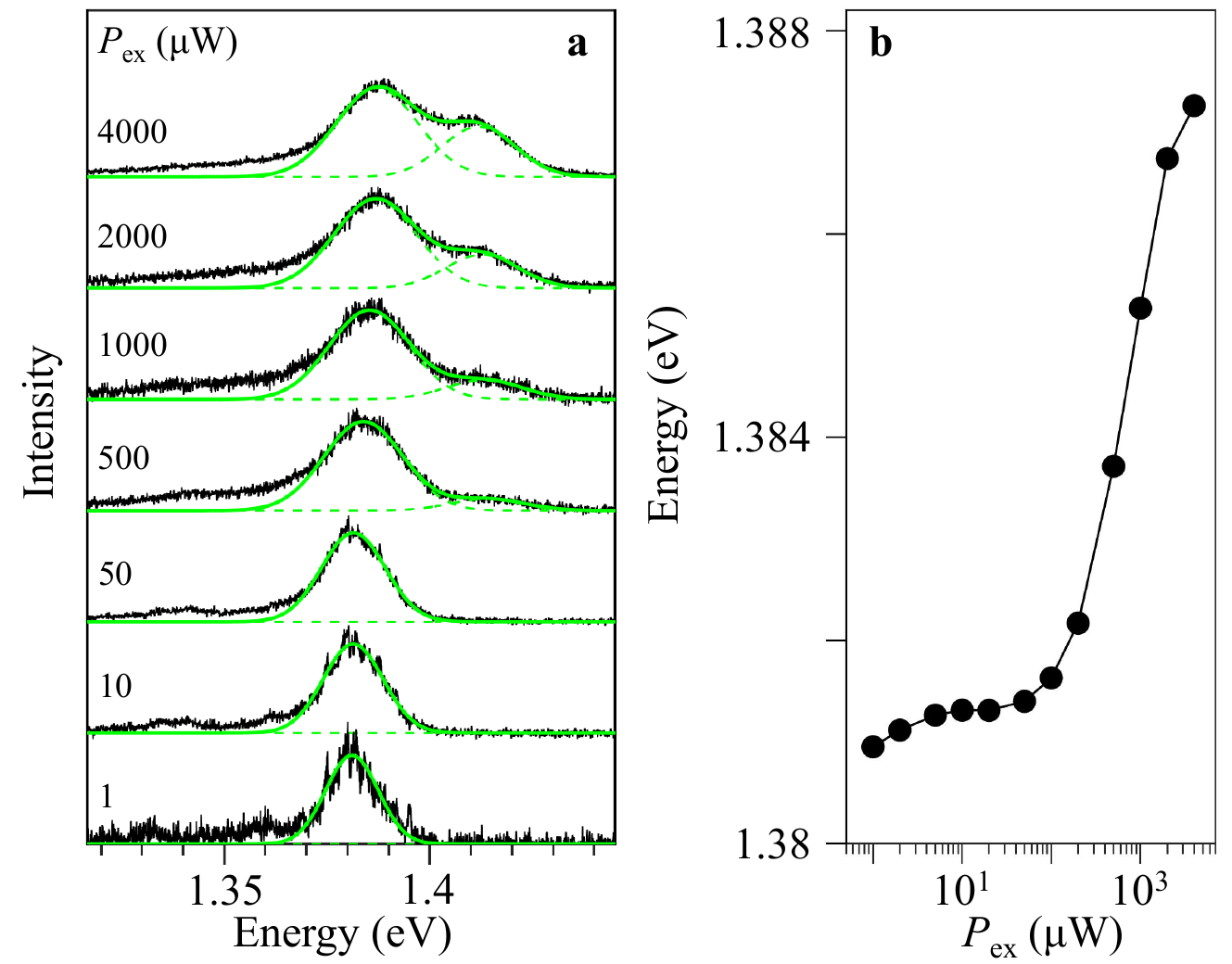}
\caption{(a) IX spectra vs. excitation power $P_{\rm ex}$. $T = 6$~K. The dashed lines show the Gaussian fits. The sum of the Gaussians is presented by the solid green line. (b) IX energy vs. $P_{\rm ex}$.
}
\end{center}
\label{fig:spectra}
\end{figure}

\vskip5mm
{\bf IX spectra}

Figure~S2 shows IX PL spectra for different excitation powers $P_{\rm ex}$. A higher-energy IX line appearing at high IX densities (Fig.~S2) was attributed to the appearance of moir{\'e} cells with double occupancy~\cite{Zhou2024}. Narrow lines appearing in the IX spectrum at lower densities (Fig.~S2) were attributed to localized states~\cite{Zhou2024}. The transport kinetics data are spectrally integrated over the entire IX PL and all IX states are included in the transport measurements. 

Shift $\delta E$ of the lower-energy IX line with $P_{\rm ex}$ is shown in Fig.~S2b. This shift is used in the paper to estimate IX density and occupancy per moir{\'e} unit cell within the mean-field model~\cite{Yoshioka1990}, as outlined in the main text. $\delta E \sim 6$~meV corresponds to $n \sim 4 \cdot 10^{11}$~cm$^{-2}$~\cite{Yoshioka1990} and $N \sim 1$ for the heterostructure. 

Figure~S2 shows that, up to the highest densities, including the Mott insulator regime, $N \sim 1$, the IX linewidth is close to the linewidth at the lowest densities. This indicates the excitonic regime for the entire IX density range in this work. The excitonic regime is consistent with the estimated IX density $n \sim 4 \cdot 10^{11}$~cm$^{-2}$ for $N \sim 1$, which is significantly lower than the density of the Mott transition to an electron-hole plasma ($n_{\rm Mott} > 10^{12}$~cm$^{-2}$)~\cite{Fogler2014, Wang2019}. In contrast to the excitonic regime, the linewidth substantially increases in the plasma regime~\cite{Wang2019}.

\begin{figure}
\begin{center}
\includegraphics[width=8.7cm]{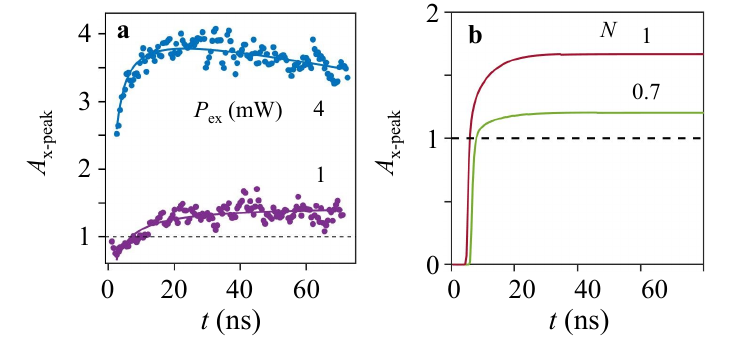}
\caption{(a) The contrast of the peak in the spatial distribution of IXs at the origin, $A_{\rm x-peak}$, vs. time for $P_{\rm ex} = 1$ and 4~mW ($N \sim 0.7$ and 1). $N$ values are estimated within the mean field model~\cite{Yoshioka1990} as outlined in the text. $T = 1.7$~K. The peak develops several ns after the excitation starts. 
(b) Simulated $A_{\rm x-peak}$ vs. time for $N = 0.7$ and 1.
}
\end{center}
\label{fig:spectra}
\end{figure}

\begin{figure}
\begin{center}
\includegraphics[width=8.7cm]{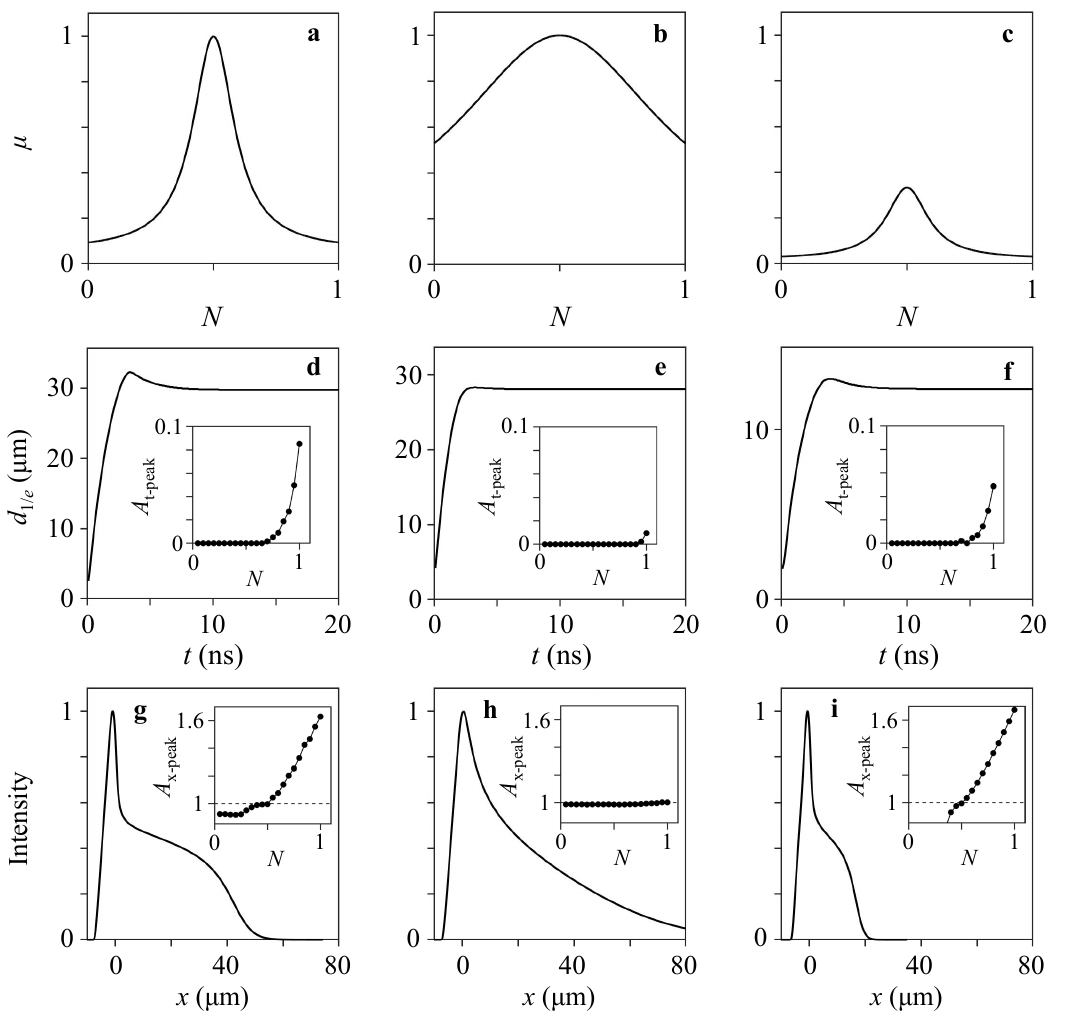}
\caption{Simulations with various $\mu(N)$. $\mu(N)$ with $\mu(1/2) = 1.7 \cdot 10^5$~cm$^2$/eVs in (a) corresponds to $\mu(N)$ in Fig.~3e. $\mu(N)$ in (b) has the same $\mu(1/2)$ and smaller contrast $\mu(1/2)/\mu(1)$. $\mu(N)$ in (c) has the same contrast $\mu(1/2)/\mu(1)$ and smaller $\mu(1/2)$. 
(d-f) $d_{1/e}(t)$ for $N = 1$.
The inserts show the contrast of the peak in $d_{1/e}(t)$.
(g-i) $I(x)$ for $N = 1$. 
The inserts show the contrast of the peak in $I(x)$ at the origin. 
$\mu(N)$ for (d-i) are shown in the corresponding column of (a-c). 
The peak in $d_{1/e}(t)$ becomes stronger as both $\mu(1/2)/\mu(1)$ and $\mu(1/2)$ increase.
}
\end{center}
\label{fig:spectra}
\end{figure}

\vskip5mm
{\bf Potential energy landscape}

Eq.~1 in the main text includes the term describing IX drift due to the in-plane potential energy landscape, $U$. This landscape arises from variations of the local IX environment in the heterostructure. Its specific origin is not essential for transport modeling by Eq.~1. $U(x)$, given by IX PL energy at different $x$, drops by $\sim 10$~meV over $\sim 10$~$\mu$m along the IX path from the excitation spot to the heterostructure edge, and is approximated by the energy gradient 1~meV/$\mu$m for the simulations with Eq.~1.

\vskip5mm
{\bf Kinetics of the peak in $I(x)$}

The emergence of non-monotonic temporal kinetics in IX transport upon entering the Mott insulator regime (Fig.~1 in the main text) is accompanied by a peak in the IX spatial distribution at the origin (Fig.~2). This peak is quantified by the non-exponentiality contrast $A_{\rm x-peak} = (I_0/I_3)/(I_3/I_6)$, where $I_0$, $I_3$, and $I_6$ is IX signal at $x = 0$, 3, and 6~$\mu$m, respectively. $A_{\rm x-peak} = 1$ for an exponential decay. For a cloud that originates at $x = 0$ and decays exponentially with increasing $x$, $A_{\rm x-peak}$ is less than 1 because of the smooth $I(x)$ maximum at the origin. A peak in IX spatial distribution, with $A_{\rm x-peak} > 1$, emerges at the origin upon entering the Mott insulator regime, $N \sim 1$, 
(Fig.~2b). This is qualitatively reproduced by simulations (Fig.~3d,g), as discussed in the main text.

In this section, we outline the kinetics of the peak in the IX spatial distribution, $I(x)$, at the origin. Figure~S3a shows that the peak at $x = 0$ develops several ns after the excitation starts. This time is comparable to the time of the peak in $d_{1/e}(t)$ (Fig.~1b). The simulations outlined in the main text also show that the peak at the origin in $I(x)$ (Fig.~S3b) and the peak in $d_{1/e}(t)$ (Fig.~3b) develop over comparable timescales. As outlined in the main text, the peak in $d_{1/e}(t)$ (Fig.~1) and the peak in $I(x)$ (Fig.~2) develop upon entering the Mott insulator regime, where IX transport is suppressed. This common origin is consistent with both peaks developing over similar timescales.

\vskip5mm
{\bf Simulations with various $\mu(N)$}

As described in the main text, the peak in $d_{1/e}(t)$ (Fig.~1) and the peak at the origin in IX spatial distribution (Fig.~2) originate from the IX transport suppression upon entering the Mott insulator regime, $N \sim 1$. This is supported by simulations of IX transport with the IX mobility falling with increasing $N$ for $1/2 < N < 1$ (Fig.~3e). These simulations show that the non-monotonic dependence of IX transport on $N$ qualitatively reproduces the observed temporal expansion and contraction of the IX cloud with the peak in $d_{1/e}(t)$ (Fig.~1), as well as the peak at $x = 0$ in $I(x)$ (Fig.~2), upon entering the Mott insulator regime. In this section, we outline how the amplitude and shape of $\mu(N)$ may affect these peaks.

Simulations with various $\mu(N)$ show that the peaks in $d_{1/e}(t)$ and $I(x)$ becomes stronger when the contrast of $\mu(N)$, quantified by $\mu(1/2)/\mu(1)$, increases (compare left and middle columns in Fig.~S4). This indicates that the emergence of the peaks in $d_{1/e}(t)$ and $I(x)$ is facilitated by a strong contrast between IX transport at $N \sim 1/2$ and $N \sim 1$. Simulations with various $\mu(N)$ also show that the peak in $d_{1/e}(t)$ becomes stronger when $\mu(1/2)$ increases (compare left and right columns in Fig.~S4). This indicates that the emergence of the peak in $d_{1/e}(t)$ is facilitated by efficient IX transport at $N \sim 1/2$. The observed peaks in $d_{1/e}(t)$ (Fig.~1) and $I(x)$ (Fig.~2) confirm that the system meets the conditions for both efficient IX transport at $N \sim 1/2$ and strong contrast in IX transport between $N \sim 1/2$ and $N \sim 1$. We note that $\mu(1/2)$ in the simulations (Figs.~3e, S4a) is high and comparable to the $\mu(1/2)$ value characterizing IX transport in the heterostructure~\cite{Zhou2026}.

\begin{figure}
\begin{center}
\includegraphics[width=4cm]{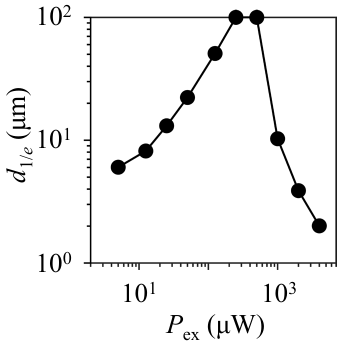}
\caption{$d_{1/e}(t)$ at long delay time $t = 70$~ns vs. $P_{\rm ex}$. $T = 1.7$~K. 
}
\end{center}
\label{fig:spectra}
\end{figure}

\vskip5mm
{\bf $d_{1/e}(t)$ at long delay times}

As outlined in the main text, at sufficiently long delay times, the IX transport distance $d_{1/e}(t)$ saturates at the value corresponding to cw laser excitation, which was considered in Ref.~\cite{Fowler-Gerace2024}. $d_{1/e}$ values at long times (Fig.~S5) are close to those in cw experiments (compare Fig.~S5 with Fig.~3a in Ref.~\cite{Fowler-Gerace2024}). However, cw experiments do not reveal how $d_{1/e}$ evolve over time to reach these values. This work shows that the $d_{1/e}$ values at long times are achieved through complex IX transport kinetics, including IX cloud expansion and contraction as the system enters the Mott insulator phase.

\vskip5mm
$^*$ These authors contributed equally: W.~J. Brunner, Zhiwen~Zhou.